\documentclass[12pt, a4paper, twoside, notitlepage]{article}

\usepackage{mathtools}
\usepackage{amsfonts,amssymb,amsthm}
\usepackage{dsfont}
\usepackage{bm}	

\usepackage[super]{nth} 

\usepackage{multirow}

\usepackage[
bookmarks=true,
bookmarksopen=true,
bookmarksnumbered=true,
pdfstartpage=1,
pdftitle={Evaluation of time series models under non-stationarity with application to the comparison of regional climate models},
pdfauthor={T. M. Erhardt, C. Czado and T. L. Thorarinsdottir},
colorlinks=true,
linkcolor=black,
anchorcolor=black,
citecolor=black,
filecolor=black,
menucolor=black,
urlcolor=black
]{hyperref}

\allowdisplaybreaks	

\newcommand{\wh}{\widehat}
\newcommand{\wt}{\widetilde}

\newcommand{\N}{\mathbb{N}}
\newcommand{\R}{\mathbb{R}}

\newcommand{\calF}{\mathcal{F}}
\newcommand{\calN}{\mathcal{N}}
\newcommand{\calC}{\mathcal{C}}
\newcommand{\calW}{\mathcal{W}}
\newcommand{\calI}{\mathcal{I}}

\newcommand{\veps}{{\varepsilon}}
\newcommand{\btheta}{\boldsymbol{\theta}}

\newcommand{\btau}{\boldsymbol{\tau}}
\newcommand{\bmu}{\boldsymbol{\mu}}
\newcommand{\bsigma}{\boldsymbol{\sigma}}

\newcommand{\bx}{{\bm{x}}}
\newcommand{\by}{{\bm{y}}}

\DeclareMathOperator{\E}{\mathbb{E}}

\DeclareMathOperator{\ind}{\mathds{1}}

\newcommand{\medianI}{\operatornamewithlimits{median}}

\DeclareMathOperator{\sSE}{SE}

\DeclareMathOperator{\sCRPS}{CRPS}

\DeclareMathOperator{\ST}{ST}
\DeclareMathOperator{\PW}{PW}
\DeclareMathOperator{\OF}{OF}
\DeclareMathOperator{\OV}{OV}
\DeclareMathOperator{\DV}{DV}

\usepackage{natbib,upgreek}

\usepackage{fancyhdr}
\usepackage{ifthen}
\begin{document}

\title{Evaluation of time series models under non-stationarity with application to the comparison of regional climate models}

\author{by\\
\\
T. M. Erhardt and C. Czado\\
Zentrum Mathematik\\
Technische Universit\"at M\"unchen\\
Parkring 13, 85748 Garching bei M\"unchen, Germany\\
\\
and\\
\\
T. L. Thorarinsdottir\\
Norwegian Computing Center\\
P.O. Box 114 Blindern, NO-0314 Oslo, Norway\\
}

\maketitle

\begin{abstract}
Different disciplines pursue the aim to develop models which characterize certain phenomena as accurately as possible. Climatology is a prime example, where the temporal evolution of the climate is modeled. In order to compare and improve different models, methodology for a fair model evaluation is indispensable. As models and forecasts of a phenomenon are usually associated with uncertainty, proper scoring rules, which are tools that account for this kind of uncertainty, are an adequate choice for model evaluation. However, under the presence of non-stationarity, such a model evaluation becomes challenging, as the characteristics of the phenomenon of interest change. We provide methodology for model evaluation in the context of non-stationary time series. Our methodology assumes stationarity of the time series in shorter moving time windows. These moving windows, which are selected based on a changepoint analysis, are used to characterize the uncertainty of the phenomenon/model for the corresponding time instances. This leads to the concept of moving scores allowing for a temporal assessment of the model performance. The merits of the proposed methodology are illustrated in a simulation and a case study.\\

\end{abstract}

\clearpage

\section[Introduction]{Introduction}\label{sec:intro}

Ever since $1990$, the Intergovernmental Panel on Climate Change (IPCC) has published regular reports on the current state of knowledge on climate change. Its latest Assessment Report \citep[AR5, see][for the synthesis report]{IPCC14} consolidates our current understanding of climate change, its causes and impacts, and discusses possible adaptation and mitigation strategies. Climate models which provide a mathematical description of certain processes in the Earth's climate system are the main tool to learn about the possible future changes of the climate.

A multitude of different climate models is used to simulate future states of the climate as described by variables such as surface temperature and precipitation, amongst others. In order to evaluate the accuracy of different climate models, simulations from these models are compared to historical observations. As \cite{ipcc2010} note, no general all-purpose metric has been found that unambiguously identifies a best model. Furthermore, different metrics may produce different model rankings \citep[e.g.][]{gleckler2008}.  

One way to compare model simulations to observations is to perform the comparison separately for each time unit (and each spatial unit). Such a point-wise comparison \citep[e.g. based on one of the metrics discussed in][]{hyndman06} is obviously the most straightforward approach to model evaluation. However, this approach is not adequate for evaluating climate models based on a daily or sub-daily temporal resolution. Climate models aim to model the long-term evolution of the climate, and can not provide accurate simulations/predictions on a daily basis. Rather, they intend to model/simulate the characteristics of the climate for longer time periods, up to several decades. In other words, the climate model output for a specific time instance is associated with uncertainty which is neglected in a point-wise evaluation.

In this setting, three questions arise:
\begin{enumerate}
	\item How can we assess the model/prediction distribution, if all we have is one time series of realizations from the model/deterministic forecasts?
	\item How can we assess the distribution of the observed quantity, if all we have is one time series of realizations of that quantity?
	\item Can we assume one and the same distribution for different time instances, that is, is the distribution stationary?
\end{enumerate}
To answer these questions we cling to our application of climate model evaluation. Under the presence of climate change and seasonality, it is obvious that we have to negate the third question. This makes it more difficult to answer the first two questions. Our answer assumes that the characteristics of the variables of interest change only gradually and can be considered as (approximately) stationary for short time windows. Then, for each of these time windows, we can construct empirical distributions based on the corresponding realizations. These empirical distributions are subsequently evaluated using the framework of proper scoring rules \citep[see e.g.][]{gneiting07} where an observed quantity is compared against a predicted distribution. 

The (moving) time windows, for which we assume stationarity, are selected using a changepoint detection algorithm \citep[pruned exact linear time (PELT) algorithm;][]{killick12}. We propose and compare three different window selection strategies. Based on the samples corresponding to these (moving) windows we compute time series of (moving) scores. This allows us to assess the model performance over time. The introduced evaluation technique is not restricted to the evaluation of climate models. Hence, we introduce it in a general setting. To learn how the approach operates in different settings we conduct a simulation study (Section \ref{sec:sim}) covering a wide range of cases which are relevant in practice. As a case study (Section \ref{sec:expl}) we evaluate daily mean surface temperature output of four regional climate models (RCMs) on a fine resolution grid covering Europe \citep[see the \emph{ENSEMBLES project}:][]{linden09}. This article and all its results are based on Chapter 5 of \citet[][]{erhardt17}.

In the \emph{ENSEMBLES project}, European climate research institutes jointly compiled a large data set of RCM simulations. One goal of this project was it to evaluate and compare the different models \citep[see e.g.][]{kjellstrom10, lorenz10}. While \citet[][]{lorenz10} compare linear trends in seasonal temperatures, \citet[][]{kjellstrom10} attempt to compare full probability distributions of temperature and precipitation against corresponding distributions from gridded observations \citep[\emph{E-OBS},][]{haylock08}. They construct empirical estimates of probability density functions by binning the data into a certain number of bins \citep{perkins2007}. As \citet[][]{kjellstrom10} note, this approach requires several subjective choices. 

Further ENSEMBLES evaluation studies include \citet[][]{landgren14} who compare local temperature and precipitation model output against different data products. They provide a model ranking based on the root mean square deviation between model output and reference data, and a measure for differences in inter-annual variability. In a study of RCM output for Canada, \citet[][]{eum12} investigate seasonal model performance based on five attributes: (i) relative absolute mean error on a daily time scale, (ii) differences in the annual variability of monthly means, (iii) differences in the spatial pattern of mean values in a certain region, (iv) discrepancy between $0.1$ and $0.9$ quantiles (``extremes'') of daily observations and model output, and (v) differences in long-term linear trends. This is an attempt to evaluate different features/moments of probabilistic output rather than considering the full distribution which comes along with the difficulty of judging the relative importance of each attribute. Alternatively, the approach proposed here considers the full distribution in time windows which are chosen such that seasonal differences are evaluated in an appropriate manner.

The remainder of the article is organized as follows. In Section \ref{sec:preliminaries} we provide the background information on proper scoring rules and the changepoint detection algorithm \citep[PELT algorithm;][]{killick12} for the specification of the moving time windows. Section \ref{sec:method} describes the new moving score methodology, whose properties are investigated in the simulation study in Section \ref{sec:sim}. Section \ref{sec:expl} provides the case study on RCMs, showing how moving scores can be applied in practice. Our conclusions are summarized in Section \ref{sec:conc}.


\section[Preliminaries]{Preliminaries}\label{sec:preliminaries}

\subsection[Proper scoring rules]{Proper scoring rules}\label{sec:proper}

To outline the theory behind proper scoring rules \citep[see][]{gneiting07} we consider the following setup: Let $\calF$ be a convex class of probability measures on a sample space $\Omega$. We consider an (observed) phenomenon with the random outcome $Y$ with (unknown) distribution $G\in\calF$ and realization $y\in\Omega$. Moreover, we consider models/predictions for $Y$ given through a random variable $X$ with (modeled) distribution $F\in\calF$ and realization $x\in\Omega$.

Then a scoring rule $s:\calF\times\Omega \rightarrow \R$ is a (negatively oriented) proper scoring rule if $S(G,G) \leq S(F,G)$ for all $F,G\in\calF$, where we define the expectation of the score $s(F,Y)$ as $S(F,G) \coloneqq \E_G \left[s(F,Y)\right]$. Hence, a scoring rule is proper if, in expectation, the random score $s(F,Y)$ is optimized if our model/prediction $F$ equals the true distribution $G$ of $Y$. Popular examples of proper scoring rules are the \emph{Squared Error (SE)}
\begin{equation}\label{eq:SEscore}
	s_{\sSE}(F,y) \coloneqq (\E_F[X]-y)^2
\end{equation}
and the \emph{Continuous Ranked Probability Score (CRPS)}
\begin{equation}\label{eq:CRPS}
	s_{\sCRPS}(F,y) \coloneqq \int_{-\infty}^{\infty}\left(F(z)-\ind\left\{z \geq y\right\}\right)^2\text{d}z  = \E_F\left|X-y\right| - \frac{1}{2}\E_F\left|X-\wt{X}\right|,
\end{equation}
where $\ind\left\{z \geq y\right\}$ equals $1$ if $z \geq y$, otherwise $0$, and $\wt{X} \sim F$ is an independent copy of $X \sim F$.

In practice, we often do not know the distribution $F$ explicitly. Instead, we have a sample $\bx \coloneqq (x_1,\ldots,x_n)$, $n\in\N$, from $F$. In that case, we consider sample versions of the scoring rules from above. The sample version of the SE in \eqref{eq:SEscore} is
\begin{equation}\label{eq:SEsample}
	s_{\sSE}(\bx, y) = \left(\frac{1}{n}\sum_{j=1}^n x_j-y\right)^2,
\end{equation}
and the sample version of the CRPS in \eqref{eq:CRPS} is
\begin{equation}\label{eq:CRPSsample}
	s_{\sCRPS}(\bx, y) = \frac{1}{n}\sum_{j=1}^n\left|x_j-y\right| - \frac{1}{2n^2} \sum_{j=1}^n\sum_{k=1}^n\left|x_j-x_k\right|.
\end{equation}

\subsection[Detection of multiple changepoints using the PELT method]{Detection of multiple changepoints using the PELT method}\label{sec:change}

To introduce the \emph{Pruned Exact Linear Time (PELT) algorithm} \citep[see][]{killick12}, we consider a time series $\by_{1:N} \coloneqq (y_1,\ldots,y_N)$. Generally speaking, a time instance $\tau\in\{1,\ldots,N-1\}$ is considered a \emph{changepoint}, if the statistical properties of the sub-series $\by_{1:\tau}=(y_1,\ldots,y_\tau)$ and $\by_{(\tau+1):N}=(y_{\tau+1},\ldots,y_N)$ differ. To explain how multiple changepoints are detected, we first introduce further notation. Let $\btau_{1:m} \coloneqq (\tau_1,\ldots,\tau_m)$
denote the ordered sequence of $m\in\{0,\ldots,N-1\}$ \emph{changepoints} of $\by_{1:N}$, where $\tau_j\in\N$, $1 \leq \tau_j \leq N-1$, $j=1,\ldots,m$. Defining $\tau_0 \coloneqq 0$ and $\tau_{m+1} \coloneqq N$, the changepoints $\btau_{1:m}$ split the sequence $\by_{1:N}$ into the $m+1$ \emph{segments} $\by_{(\tau_j+1):\tau_{j+1}}$ with \emph{segment lengths} $(\tau_{j+1}-\tau_j)$, $j=0,\ldots,m$. We call $\btau_{0:(m+1)} \coloneqq (\tau_0,\btau_{1:m},\tau_{m+1})$ an \emph{$(m+1)$-segmentation} of $\by_{1:N}$.

The PELT algorithm detects multiple changepoints by minimizing a \emph{target function}
\begin{equation}\label{eq:changecost}
	\sum_{j=0}^{m}\calC(\by_{(\tau_j+1):\tau_{j+1}}) + \kappa m.
\end{equation}
The target function is a sum of \emph{cost functions} $\calC:\R^n \to \R$, $n\in\N$, assigning a cost to the segment $\by_{(\tau_j+1):\tau_{j+1}}$ for all $j=0,\ldots,m$, and a \emph{penalty term} $\kappa m$ which is supposed to prevent overfitting by penalizing the number of changepoints $m$.

As we are particularly interested in detecting changes in mean and variance, we assume that the time series observations $y_1,\ldots,y_N$ come from a normal distribution $\calN(\mu_j,\sigma_j^2)$, where either the (unknown) mean $\mu_j$ and/or the (unknown) variance $\sigma_j^2$ change after certain (unknown) time instances (changepoints) $\btau_{1:m}$. In this case, we can select the cost function $\calC$ as twice the negative log-likelihood corresponding to a normal distribution, where the unknown parameters $\mu_j$ and $\sigma_j^2$ are replaced by their maximum likelihood estimators
\begin{equation*}
	\wh{\mu}_j = \frac{1}{\tau_{j+1}-\tau_j}\sum_{i=\tau_j+1}^{\tau_{j+1}}y_i \qquad\text{and}\qquad \wh{\sigma}_j^2 = \frac{1}{\tau_{j+1}-\tau_j}\sum_{i=\tau_j+1}^{\tau_{j+1}} \left(y_i - \wh{\mu}_j\right)^2,
\end{equation*}
respectively. Hence, the cost $\calC(\by_{(\tau_j+1):\tau_{j+1}})$ of a segment $\by_{(\tau_j+1):\tau_{j+1}}$ of an arbitrary $(m+1)$-segmentation $\btau_{0:(m+1)}$ of the ordered sequence $\by_{1:N}$ is given by
\begin{equation}\label{eq:cost}
	\calC(\by_{(\tau_j+1):\tau_{j+1}}) \coloneqq (\tau_{j+1}-\tau_j) \left\{\ln\left[\frac{2\pi}{\tau_{j+1}-\tau_j}\sum_{i=\tau_j+1}^{\tau_{j+1}}\left(y_i - \frac{\sum_{k=\tau_j+1}^{\tau_{j+1}}y_k}{\tau_{j+1}-\tau_j}\right)^2\right] + 1\right\}.
\end{equation}
Note, that the estimation of the variance demands a minimum segment length $(\tau_{j+1}-\tau_j)$ of $2$ for all segments $\by_{(\tau_j+1):\tau_{j+1}}$, $j=0,\ldots,m$.
Common choices for $\kappa$ are $\kappa=2p$ (cp. Akaike information criterion) or $\kappa=p\ln(N)$ (cp. Bayesian information criterion), which depend on the number $p$ of additional parameters needed per additional segment.

The PELT algorithm itself builds on the \emph{Optimal Partitioning (OP)} algorithm introduced by \citet{jackson05}. It minimizes \eqref{eq:changecost} based on a recursive detection of the changepoints of the sub-series $\by_{1:s}$, $s \leq N$. \citet{killick12} found that the performance of the OP algorithm can be improved by reducing the set of candidate changepoints in the recursion (so called pruning), in order to avoid irrelevant computations.

\section[Methodology]{Methodology}\label{sec:method}

In this section we propose a methodology for the (proper) evaluation (see Section \ref{sec:proper}) of time series models in a non-stationary context. In comparison to a purely point-wise evaluation where a model output $x_t$ at time point $t$ is compared against the corresponding observation $y_t$, our approach also considers higher order structures of the time series. Subjective choices regarding the importance of different model characteristics are avoided. As the approach is designed to deal with seasonality, it does not require a separate consideration of different selected seasons.

In the following, we consider the following setup: Let $\calF$ be a convex class of probability measures on a sample space $\Omega$.
We consider an (observed) \emph{phenomenon} with the random outcomes $Y_t$ with (unknown) distributions $G_t\in\calF$ and realizations $y_t\in\Omega$ ($t=1,\ldots,N$). 
Further, we consider a \emph{model/prediction} for $Y_t$, $t=1,\ldots,N$, given by random variables $X_t$ with (modeled) distributions $F_t\in\calF$ and realizations $x_t\in\Omega$ ($t=1,\ldots,N$).

\subsection[Discussion of naive evaluation approaches]{Discussion of naive evaluation approaches}\label{sec:naive}

\paragraph{Evaluation under a stationarity (ST) assumption}
Let us first assume that $Y_t$ and $X_t$, $t=1,\ldots,N$, are stationary. Hence, $G_t=G$ and $F_t=F$ for all $t=1,\ldots,N$. Under this assumption, we can evaluate the model $X_t$ for $Y_t$, using scores $s^{\ST}(F,y_t)$, $t=1,\ldots,N$. To reflect that they are calculated based on a stationarity assumption, we call them \emph{ST-scores}.
Their corresponding sample version is given by $s^{\ST}(\bx,y_t)$, $t=1,\ldots,N$, where $\bx=(x_1,\ldots,x_N)$. In practice the stationarity assumption is violated in most cases. Hence, ST-score based evaluation is inappropriate. A possible exception are very short time series data, such as annual series, where non-stationarity cannot be estimated reliably. 

\paragraph{Point-wise (PW) evaluation}
Let us now consider the non-stationary case. Then, it generally holds that $G_s \neq G_t$ and $F_s \neq F_t$ for $s \neq t$. A naive approach for model evaluation is to use sample scores $s^{\PW}(x_t,y_t)$, $t=1,\ldots,N$. As they compare the realizations of $Y_t$ and $X_t$ point-wise (separately for each $t=1,\ldots,N$), we call them \emph{PW-scores}. From Equations \eqref{eq:SEsample} and \eqref{eq:CRPSsample} we obtain $s_{\sSE}^{\PW}(x_t,y_t) = \left(x_t - y_t\right)^2$ and
$s_{\sCRPS}^{\PW}(x_t,y_t) = \left|x_t - y_t\right|$ for all $t=1,\ldots,N$. For a comprehensive model evaluation, PW-scores are also not satisfying, since they do not account for higher order structures of the observed/modeled phenomenon. They evaluate only how similar the phenomenon and the model/prediction behave in terms of their (temporarily varying) mean. Such point-wise model evaluation completely neglects differences in higher order moments (e.g. variance). A PW-score treats a discrepancy in $x_t$ and $y_t$ that occurs due to a falsely specified model/prediction mean in the same way as a discrepancy that occurs due to a high uncertainty of the phenomenon.

\subsection[Moving scores]{Moving scores}\label{sec:moving}

To deal with non-stationarity and to consider higher order structures of the observed phenomenon in the evaluation we suggest the following. We assume that the phenomenon $Y_t$ and the corresponding model/prediction $X_t$ are (approximately) stationary for short time intervals. Then, for each time instance $t=1,\ldots,N$ (\emph{window location}), we select integers $\delta_t^-,\delta_t^+ \in \{0,\ldots,N-1\}$ which determine the width of a \emph{moving (time) window}
\begin{equation*}
	\calW(t) = \{t-\delta_t^-,\ldots,t,\ldots,t+\delta_t^+\} \subset\{1,\ldots,N\},
\end{equation*}
such that it completely lies within the observation period and such that $Y_s$ and $X_s$ are (approximately) stationary for all $s\in\calW(t)$. Thus, we assume that for $s\in\calW(t)$, $Y_s$ and $X_s$ are distributed according to distributions $G_{t}$ and $F_{t}$ (depending on the window location $t$), respectively. Hence, in order to evaluate the model/prediction $X_t$ for the phenomenon $Y_t$, for a specific time instance $t=1,\ldots,N$, we substitute the \emph{theoretical scores} $s(F_t,y_t)$ by sample scores
\begin{equation}\label{eq:mscore}
	s(\bx_{\calW(t)},y_t)
\end{equation}
based on the sub-samples $\bx_{\calW(t)} \coloneqq \{x_s: s\in\calW(t)\}$.
Since the scores \eqref{eq:mscore}  are based on moving time windows, we call them \emph{moving scores}.
In contrast to the theoretical scores, the moving scores \eqref{eq:mscore} are \emph{empirical scores}, since they are sample-based. Same holds for the PW- and ST-scores introduced in Section \ref{sec:naive}. In order to rank different models, we can consider the score averages over $t=1,\ldots,N$.

Bearing in mind the discussion of Section \ref{sec:naive} it becomes clear that the moving score methodology is a compromise between a point-wise evaluation (which does not account for higher order structures) and an evaluation which ignores non-stationarity. There is a trade-off between small and large moving windows $\calW(t)$: In order to not violate the \emph{stationarity assumption}, we have to keep the moving windows small enough. Considering the formula \eqref{eq:CRPSsample} for the computation of sample CRPS, we see that its usage in a moving window based evaluation results in a \emph{computation time} which grows quadratically with increasing window width $|\calW(t)|$. Hence, if the full data set consists of multiple long time series for several models (and possibly many different spatial locations), a moving window based evaluation may become infeasible if the moving windows are too large\footnote{For such cases, see \citet{hersbach2000} who proposes an equivalent yet more computationally efficient computation of the sample version of the CRPS.}. However, if the moving windows are too small (small sample size), the samples $\bx_{\calW(t)}$ coming from these small windows $\calW(t)$ can only achieve an \emph{inaccurate approximation} of the true distributions $F_t$.
We see that there are reasons supporting both small and large sizes of the moving windows $\calW(t)$. In the subsequent section we introduce three different window selection strategies, all of which make a compromise between small and large windows.

\subsection[Selection of moving windows]{Selection of moving windows}\label{sec:windows}

Since the aim of the model evaluation is to compare several models against the same time series of realizations $y_1,\ldots,y_N$, it is meaningful to determine the moving windows $\calW(t)\subset\{1,\ldots,N\}$, $t=1,\ldots,N$ once based on the realized time series. Below we introduce three alternative window selection approaches based on the changepoint analysis described in Section \ref{sec:change}. In a first step, $m < N-1$ changepoints $\btau_{1:m}=(\tau_1,\ldots,\tau_m)$ of $y_1,\ldots,y_N$ are detected using the PELT algorithm with the cost function in Equation \eqref{eq:cost} which assumes varying means and variances for different segments of the time series. Hence, it allows to detect changes in mean and/or variance. For the penalty constant $\kappa$ we choose $\kappa=p\ln(N)$ with $p=3$, since we count three additional parameters (mean, variance, changepoint) per segment. Moreover, we demand that the minimum segment length is $11$ (see our discussion in Section \ref{sec:moving}), that is it must hold $(\tau_{j+1}-\tau_j)>10$, $j=0,\ldots,m$. In a second step, (different types of) moving windows are specified, based on the selected $(m+1)$-segmentation $\btau_{0:(m+1)} = (0,\btau_{1:m},N)$:

\paragraph{Overlapping windows with fixed width (OF)}
For the first approach, we consider moving time windows $\calW^{\OF}(t)$ of a fixed width $\omega^{\OF}\coloneqq\left|\calW^{\OF}(t)\right|$ except for edge effects at both ends of the time interval $\{1,\ldots,N\}$. We center the overlapping (symmetric) windows $\calW^{\OF}(t)$ around their window locations $t$. To obtain $\calW^{\OF}(t)$, we first calculate the \emph{median segment length}
$\lambda \coloneqq \medianI_{j=0,\ldots,m}\left(\tau_{j+1}-\tau_j\right)$,
where $\medianI_{\calI}\left(\cdot\right)$ denotes the sample median of quantities indexed by the index set $\calI$. Then, we compute the \emph{fixed window width parameter} defined as
$\delta^{\OF} \coloneqq \left\lfloor(\lambda-1)/2\right\rfloor$,
where $\left\lfloor\cdot\right\rfloor$ rounds a number to its next smaller integer. Defining the window width parameter $\delta_t^{\OF}$ as
\begin{equation*}
	\delta_t^{\OF} \coloneqq
	\begin{cases}
    t-1, & \text{for } t=1,\ldots,\delta^{\OF}, \\
    \delta^{\OF}, & \text{for } t=1+\delta^{\OF},\ldots,N-\delta^{\OF}, \\
    N-t, & \text{for } t=N-\delta^{\OF}+1,\ldots,N,
  \end{cases}
\end{equation*}
we ensure that the (moving) \emph{overlapping windows with fixed width} (\emph{OF-windows}) defined as
\begin{equation}\label{eq:OFwindows}
	\calW^{\OF}(t) \coloneqq \left\{t-\delta_t^{\OF},\ldots,t,\ldots,t+\delta_t^{\OF}\right\},
\end{equation}
are symmetric and $\calW^{\OF}(t)\subset\{1,\ldots,N\}$ for all $t=1,\ldots,N$. The corresponding window widths equal $\omega_t^{\OF}=2\delta_t^{\OF}+1$. The windows $\calW^{\OF}(t)$ for $t=1,\ldots,\delta^{\OF}$ and $t=N-\delta^{\OF}+1,\ldots,N$ are the \emph{border cases}. Moving scores obtained based on OF-windows will be referred to as \emph{OF-scores}.

\begin{figure}[htbp]
	\centering
		\includegraphics[width=0.93\textwidth]{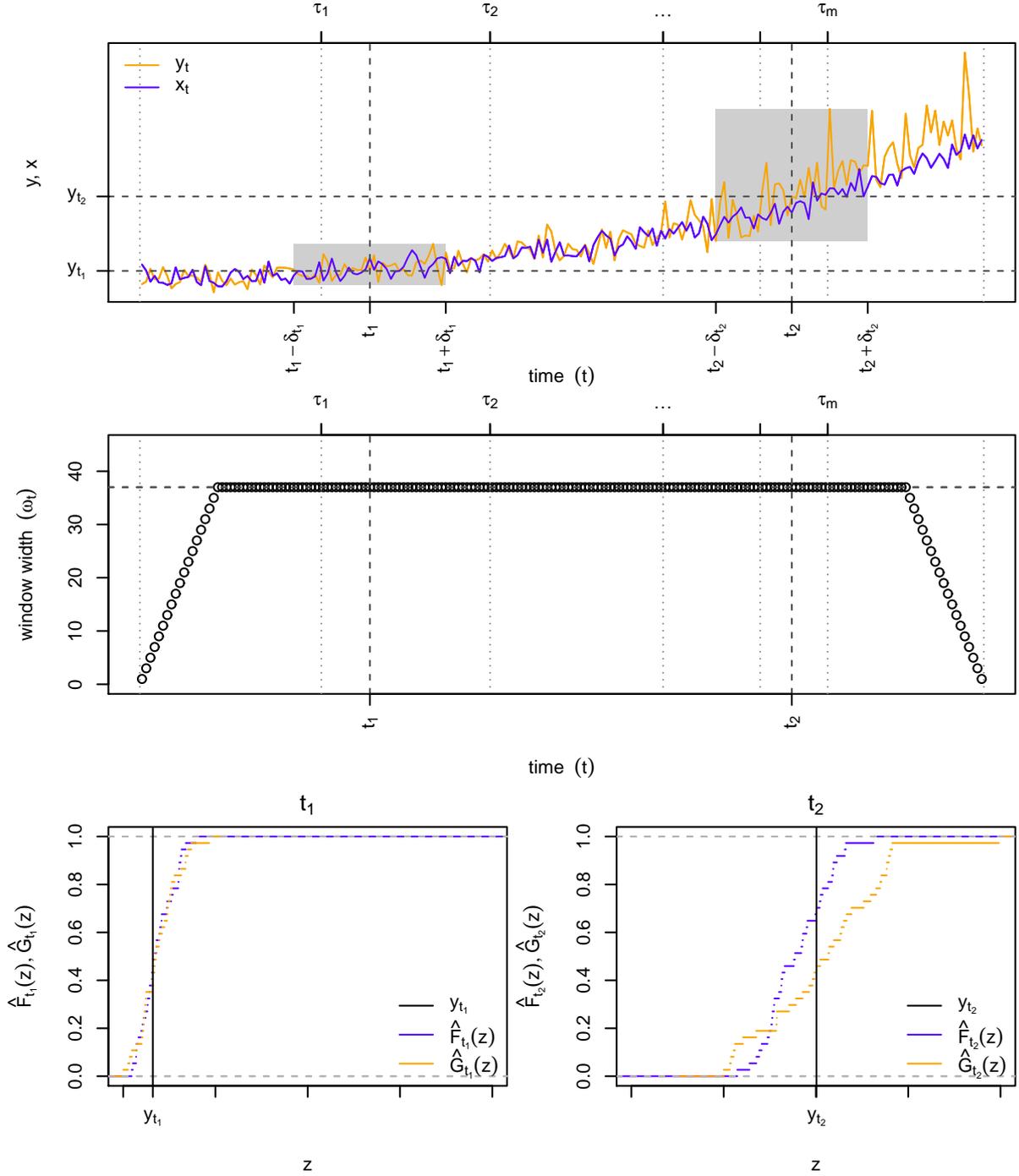}
	\caption{Illustration of moving window methodology for overlapping windows with fixed width (OF): Realizations $y_t$ (orange) and $x_t$ (blue), $t=1,\ldots,N$ (of phenomenon $Y_t$ and corresponding model/prediction $X_t$), detected changepoints $\btau_{1:m}$, and moving windows $\calW^{\OF}(t_1)$ and $\calW^{\OF}(t_2)$ (gray) for two selected time instances $t_1$ and $t_2$ (upper panel). Selected window width (middle panel). Empirical CDFs for phenomenon ($\wh{G}_{t}$) and model/prediction ($\wh{F}_{t}$) based on window $\calW^{\OF}(t)$, and observation ($y_t$) in the window location ($t$), for the time instances $t_1$ (lower left panel) and $t_2$ (lower right panel).}
	\label{fig:OF}
\end{figure}

The top panel of Figure \ref{fig:OF} illustrates two time series ($y_t$ and $x_t$, $t=1,\ldots,N$) of realizations from a phenomenon $Y_t$ and a corresponding model/prediction $X_t$. Moreover, it shows which changepoints $\btau_{1:m}$ were detected by the PELT algorithm. The middle panel shows which window width was selected by the OF-method for each time instance $t=1,\ldots,N$. The corresponding moving windows $\calW^{\OF}(t_1)$ and $\calW^{\OF}(t_2)$ for two selected time instances $t_1$ and $t_2$ are illustrated as gray boxes in the upper panel. The lower panel shows for both OF-windows (located at $t_1$ and $t_2$), how the empirical CDFs for both the phenomenon ($\wh{G}_{t}$) and the model/prediction ($\wh{F}_{t}$) differ from each other and how much they deviate from the observation ($y_t$) in the window location ($t$).

\paragraph{Overlapping windows with varying width (OV)}
For the second approach, we consider moving time windows $\calW^{\OV}(t)$ with varying width $\omega_t^{\OV}\coloneqq\left|\calW^{\OV}(t)\right|$. Again, we center the overlapping windows around their window location $t$. To obtain the varying windows $\calW^{\OV}(t)$, we first determine the centers of the segments $\{\tau_j+1,\ldots,\tau_{j+1}\}$, $j=0,\ldots,m$, of the $(m+1)$-segmentation $\btau_{0:(m+1)}$ defined as $\gamma_j \coloneqq (\tau_j+1+\tau_{j+1})/2$, $j=0,\ldots,m$.
Then, we linearly interpolate the corresponding segment lengths $\lambda_j \coloneqq \left(\tau_{j+1}-\tau_j\right)$ between the segment centers $\gamma_j$, $j=0,\ldots,m$. Hence, the \emph{interpolated segment lengths} are given by
$\varsigma(t) \coloneqq [(\gamma_{j+1}-t)\lambda_j + (t-\gamma_j)\lambda_{j+1}]/(\gamma_{j+1}-\gamma_j)$, for $t\in[\gamma_j,\gamma_{j+1}]$, $j=0,\ldots,m$.
Then, the (symmetric) \emph{overlapping windows with varying width (OV-windows)} are defined by
\begin{equation}\label{eq:OVwindows}
	\calW^{\OV}(t) \coloneqq \left\{t-\delta_t^{\OV},\ldots,t,\ldots,t+\delta_t^{\OV}\right\},
\end{equation}
where
\begin{equation*}
	\delta_t^{\OV} \coloneqq
	\begin{cases}
    t-1, & \text{for } t=1,\ldots,\left\lfloor\gamma_0\right\rfloor, \\
    \left\lfloor\left(\varsigma(t)-1\right)/2\right\rfloor, & \text{for } t=\left\lceil\gamma_0\right\rceil,\ldots,\left\lfloor\gamma_{m+1}\right\rfloor, \\
    N-t, & \text{for } t=\left\lceil\gamma_{m+1}\right\rceil,\ldots,N.
  \end{cases}
\end{equation*}
Here, $\left\lfloor\cdot\right\rfloor$ and $\left\lceil\cdot\right\rceil$ round a number to its closest smaller and larger integer, respectively. Note that for the \emph{border cases} ($t=1,\ldots,\left\lfloor\gamma_0\right\rfloor$ and $t=\left\lceil\gamma_{m+1}\right\rceil,\ldots,N$) the moving windows are again defined such that they completely lie within $\{1,\ldots,N\}$. The varying window widths equal $\omega_t^{\OV}=2\delta_t^{\OV}+1$. Moving scores obtained based on OV-windows will be referred to as \emph{OV-scores}. An illustration of the OV-method in analogy to Figure \ref{fig:OF} is provided by Figure S.2 in the supplementary material.

\paragraph{Disjoint windows with varying width (DV)}
For the third approach, we consider disjoint moving time windows $\calW^{\DV}(t)$ with varying width $\omega_t^{\DV}\coloneqq\left|\calW^{\DV}(t)\right|$ given by the segments of the $(m+1)$-segmentation $\btau_{0:(m+1)}$. Hence, the \emph{disjoint windows with varying width (DV-windows)} are defined by
\begin{equation}\label{eq:DVwindows}
	\calW^{\DV}(t) \coloneqq \left\{\tau_j+1,\ldots,\tau_{j+1}\right\}, \quad\text{for } t=\tau_j+1,\ldots,\tau_{j+1},\,j=0,\ldots,m.
\end{equation}
The varying window widths equal the corresponding segment lengths, $\omega_t^{\DV} \coloneqq \left(\tau_{j+1}-\tau_j\right)$, for $t=\tau_j+1,\ldots,\tau_{j+1}$, $j=0,\ldots,m$. Moving scores based on DV-windows will be referred to as \emph{DV-scores}. An illustration of the DV-method in analogy to Figure \ref{fig:OF} is provided by Figure S.3 in the supplementary material.


\section[Simulation study]{Simulation study}\label{sec:sim}

\subsection[Organization of simulation study]{Organization of simulation study}\label{sec:simorg}

To judge the applicability of moving scores based on different moving window selection approaches, we conduct a simulation study under three different scenarios. We first consider a \emph{changepoint scenario} (C, Section \ref{sec:cpts}),
where the characteristics (mean and variance) of the phenomenon of interest change at certain (unknown) time instances. Moreover, with climate model evaluation in mind, we consider a \emph{trend scenario} (T, Section \ref{sec:trend}) and a \emph{periodicity scenario} (P, Section \ref{sec:periodicity}).

\paragraph{General setup}
For each scenario, we consider a \emph{phenomenon} $Y_t$ and \emph{five models} $X_t^k$ for that phenomenon ($t=1,\ldots,N$), numbered by $k=1,\ldots,5$. We assume that $Y_t$, $X_t^k$, $k=1,\ldots,5$, are normally distributed for all $t=1,\ldots,N$, that is
\begin{align}
	Y_t & \sim \calN\left(\mu_{0,t},\sigma_{0,t}^2\right) \quad \text{and} \label{eq:phenomenon}\\
	X_t^k & \sim \calN\left(\mu_{k,t},\sigma_{k,t}^2\right),\quad k=1,\ldots,5. \label{eq:models}
\end{align}
Since we simulate the phenomenon, we also refer to $Y_t$, $t=1,\ldots,N$, as the \emph{data generating process (DGP)}.
Each result of the simulation study is based on $10{,}000$ replications of the data generating process \eqref{eq:phenomenon} and the corresponding model \eqref{eq:models}.

\paragraph{Methods}
For each replication, we compute time series of moving scores \eqref{eq:mscore} for the different window selection approaches dicussed in Section \ref{sec:windows}.  Moreover, we also compute point-wise (PW) scores. In case of the changepoint (C) and the trend scenario (T), we also consider ST-scores which assume stationarity of the time series. To get an idea of the temporal evolution of these (empirical) scores, we average the obtained score time series over all $10{,}000$ replications. Furthermore, we compare them to the corresponding (averaged) time series of theoretical (Theo.) scores, which can be obtained, since we know the distributions of the data generating process \eqref{eq:phenomenon} and each of the five models \eqref{eq:models}.

\subsection{Changepoint scenario}\label{sec:cpts}

In order to describe the changepoint scenario (C) we consider time varying means and standard deviations which change after certain changepoints $\btau_{1:m}\in\R^{m}$ (see Section \ref{sec:change}). Hence, the time varying means and standard deviations are defined as
$M_\text{C}(t;\btau_{0:(m+1)},\bmu) \coloneqq \mu_j$ and
$S_\text{C}(t;\btau_{0:(m+1)},\bsigma) \coloneqq \sigma_j$, for $t\in\{\tau_j+1,\ldots,\tau_{j+1}\}$
and parameter vectors $\bmu=(\mu_0,\ldots,\mu_m)\in\R^{m+1}$ and $\bsigma=(\sigma_0,\ldots,\sigma_m)\in\R^{m+1}$, respectively.

\paragraph{Changepoint scenario (C)}
The subsequent equations define the data generating process (C0) and the five models (C1)--(C5), where we consider time series of length $N=200$ with $m=2$ changepoints and corresponding 3-segmentation $\btau_{0:3}=(0,80,130,200)$.

\begin{alignat*}{4}
	&	\mbox{(C0)}	&& \quad Y_t && \sim\calN(M_\text{C}(t;\btau_{0:3},\bmu^0),[S_\text{C}(t;\btau_{0:(m+1)},\bsigma^0)]^2)	&& \quad \mbox{(data generating process),}	\\
	\hline
	&	\mbox{(C1)}	&& \quad X_t^1 && \sim\calN(M_\text{C}(t;\btau_{0:3},\bmu^1),[S_\text{C}(t;\btau_{0:(m+1)},\bsigma^1)]^2)	&& \quad \mbox{(true model),}	\\
	&	\mbox{(C2)}	&& \quad X_t^2 && \sim\calN(M_\text{C}(t;\btau_{0:3},\bmu^2),[S_\text{C}(t;\btau_{0:(m+1)},\bsigma^2)]^2)	&& \quad \mbox{(constant mean),}	\\
	&	\mbox{(C3)}	&& \quad X_t^3 && \sim\calN(M_\text{C}(t;\btau_{0:3},\bmu^3),[S_\text{C}(t;\btau_{0:(m+1)},\bsigma^3)]^2)	&& \quad \mbox{(constant variance),}	\\
	&	\mbox{(C4)}	&& \quad X_t^4 && \sim\calN(M_\text{C}(t;\btau_{0:3},\bmu^4),[S_\text{C}(t;\btau_{0:(m+1)},\bsigma^4)]^2)	&& \quad \mbox{(constant mean and variance),}	\\
	&	\mbox{(C5)}	&& \quad X_t^5 && \sim\calN(M_\text{C}(t;\btau_{0:3},\bmu^5),[S_\text{C}(t;\btau_{0:(m+1)},\bsigma^5)]^2)	&& \quad \mbox{(wrong mean and variance),}
\end{alignat*}
\begin{alignat*}{4}
	&	\mbox{(C0)}	&& \quad \bmu^0 = (0,1,0),	&& \quad \bsigma^0 = (0.9,0.9,0.3),	&& \quad \mbox{(data generating process),}	\\
	\hline
	&	\mbox{(C1)}	&& \quad \bmu^1 = \bmu^0,	&& \quad \bsigma^1 = \bsigma^0,	&& \quad \mbox{(true model),}	\\
	&	\mbox{(C2)}	&& \quad \bmu^2 = (0.25,0.25,0.25),	&& \quad \bsigma^2 = \bsigma^0,	&& \quad \mbox{(constant mean),}	\\
	&	\mbox{(C3)}	&& \quad \bmu^3 = \bmu^0,	&& \quad \bsigma^3 = (0.6,0.6,0.6),	&& \quad \mbox{(constant variance),}	\\
	&	\mbox{(C4)}	&& \quad \bmu^4 = \bmu^2,	&& \quad \bsigma^4 = \bsigma^3,	&& \quad \mbox{(constant mean and variance),}	\\
	&	\mbox{(C5)}	&& \quad \bmu^5 = (0.1,0.9,0.1),	&& \quad \bsigma^5 = \bsigma^3,	&& \quad \mbox{(wrong mean and variance).}
\end{alignat*}
Whereas the true model (C1) is equivalent to the DGP (C0) in terms of its parametrization, models (C2)--(C5) differ from (C0) in at least one parameter.
Model (C2) assumes a constant mean, but captures the change in the variance. While model (C3) models the change in the mean correctly, it assumes a constant variance. Models (C4) and (C5) are misspecified both in terms of mean and variance with the misspecification of the mean being less severe for (C5). Figure S.4 in the supplementary material illustrates one replication of the DGP (C0) and the corresponding selection of moving windows. The PELT algorithm finds two changepoints in nearly all the simulated series with an estimated fixed window width of $69$ or $71$ time points for most cases. 

\begin{figure}[hbp]
	\centering
		\includegraphics[width=1.00\textwidth]{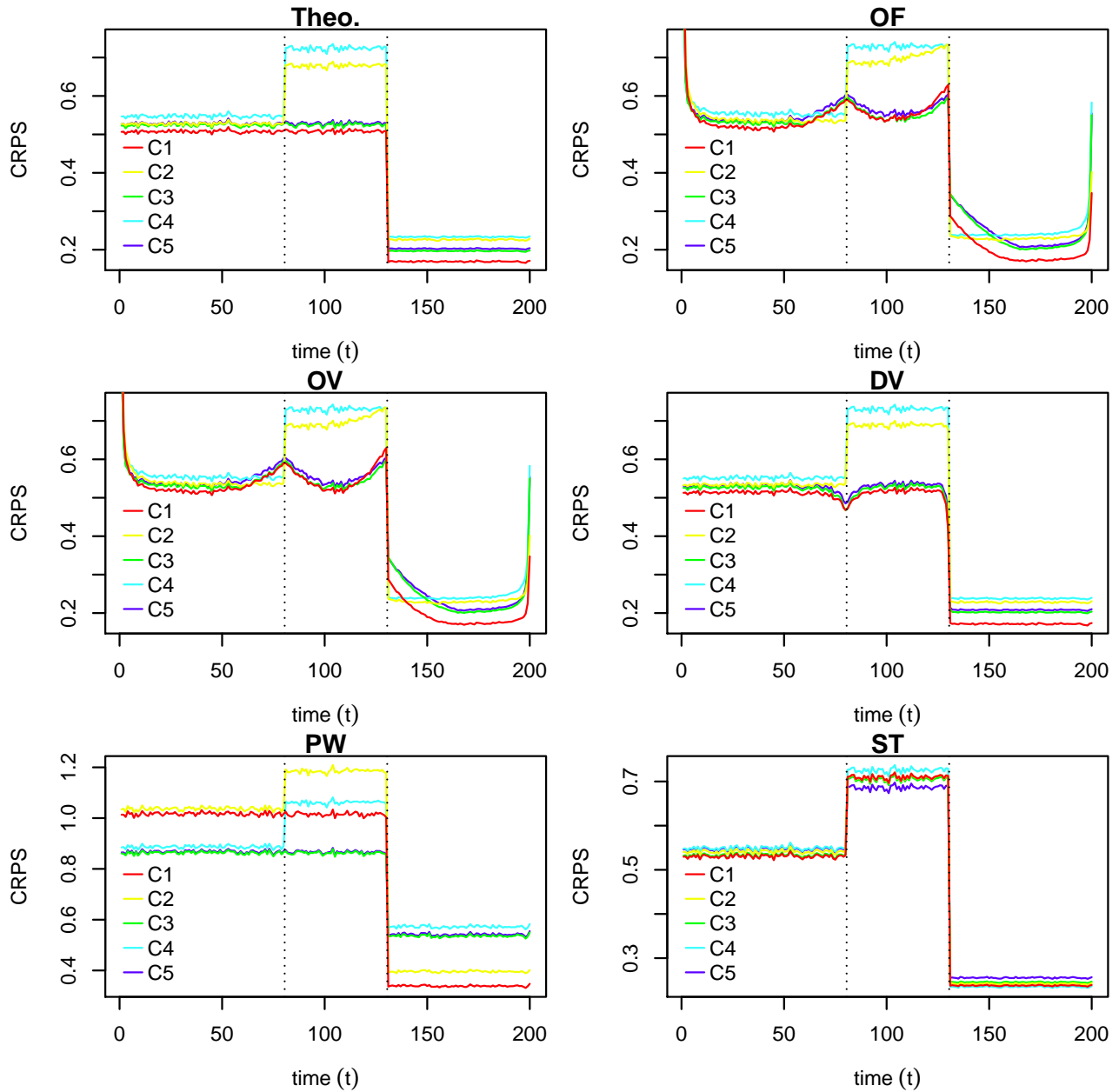}
	\caption{\textbf{Changepoint scenario (C):} Comparison of models (C1)--(C5) based on (time series of) theoretical/empirical CRPS. Comparison based on theoretical scores (upper left panel), moving scores computed using the OF approach (upper right panel), moving scores computed using the OV approach (middle left panel), moving scores computed using the DV approach (middle right panel), point-wise (PW) scores (lower left panel) and based on ST-scores under the assumption of stationarity (lower right panel).}
	\label{fig:Ccrps}
\end{figure}

\paragraph{Time series of (moving) scores}
Figure \ref{fig:Ccrps} compares models (C1)--(C5) based on time series of theoretical/empirical CRPS. We observe that the temporal evolution of the PW- and the ST-scores differs considerably from their theoretical counterparts. The observed relative model rankings are wrong for all time instances. We further observe that the moving OF- and OV-scores provide faulty model rankings for time instances close to the changepoints. The DV-scores mirror the temporal evolution of the theoretical scores best and mostly provide the correct model rankings.

\paragraph{Model rankings}
Table \ref{tab:Crank} shows which evaluation approaches rank the models (C1)--(C5) correctly and which do not. It provides average SE and CRPS over time instances and replications for the different evaluation approaches (OF, OV, DV, PW and ST) and compares them to the corresponding theoretical (Th) values. The corresponding model rankings under each method are given in the right half of the table. If the score averages of two or more models are equal, all of these models get the same (minimum) rank. We expect the true model (C1) to be ranked lowest (best), and the ranking should be identical to that based on the theoretical (Th) scores. These two criteria are fulfilled for the CRPS calculated based on the OF, OV and DV approach. The SE score is unable to differentiate between all five models as its evaluation is based on the mean value only. In particular, the empirical evaluation approaches rank model (C3) higher than the true model (C1). In the theoretical SE in \eqref{eq:SEscore}, the realization is evaluated against the expected value of the model/prediction. As model (C3) has the correct mean but a smaller variance than the true model (C1), the realized values under (C3) are likely to be closer to the true mean of the data generating process. The PW- and the ST-approach yield erroneous model rankings under both scores.

\begin{table}[t]
\caption{\label{tab:Crank}\textbf{Changepoint scenario (C):} Comparison/ranking of models (C1)--(C5) based on average (empirical/theoretical) SE and CRPS. Average scores (left) and corresponding model rankings (right) are provided, distinguishing between theoretical (Th) and empirical scores computed using different approaches (OF, OV, DV, PW, ST). The rankings based on the theoretical scores are considered the true model rankings used to judge the rankings given by the empirical scores.}
\begin{center}
\fbox{
\begin{tabular}{ll|c|ccccc|c|ccccc}
	&	&	\multicolumn{6}{c|}{average scores}	&	\multicolumn{6}{c}{model rank} \\ 
	& & Th & OF & OV & DV & PW & ST & Th & OF & OV & DV & PW & ST \\ 
  \hline
\multirow{5}{*}{\rotatebox[origin=c]{90}{SE}}
	& (C1) & 0.558 & 0.637 & 0.627 & 0.559 & 1.117 & 0.748 & 1 & 2 & 2 & 2 & 4 & 3 \\ 
  & (C2) & 0.746 & 0.764 & 0.765 & 0.754 & 1.305 & 0.748 & 4 & 5 & 5 & 5 & 5 & 3 \\ 
  & (C3) & 0.558 & 0.632 & 0.622 & 0.556 & 0.919 & 0.747 & 1 & 1 & 1 & 1 & 1 & 1 \\ 
  & (C4) & 0.746 & 0.759 & 0.759 & 0.751 & 1.107 & 0.747 & 4 & 4 & 4 & 4 & 3 & 1 \\ 
  & (C5) & 0.568 & 0.648 & 0.638 & 0.570 & 0.929 & 0.750 & 3 & 3 & 3 & 3 & 2 & 5 \\ 
	\hline
\multirow{5}{*}{\rotatebox[origin=c]{90}{CRPS}}
  & (C1) & 0.389 & 0.425 & 0.422 & 0.392 & 0.779 & 0.473 & 1 & 1 & 1 & 1 & 3 & 1 \\ 
  & (C2) & 0.460 & 0.476 & 0.476 & 0.466 & 0.850 & 0.479 & 4 & 4 & 4 & 4 & 5 & 3 \\ 
  & (C3) & 0.410 & 0.441 & 0.439 & 0.411 & 0.749 & 0.475 & 2 & 2 & 2 & 2 & 1 & 2 \\ 
  & (C4) & 0.482 & 0.494 & 0.495 & 0.487 & 0.821 & 0.483 & 5 & 5 & 5 & 5 & 4 & 5 \\ 
  & (C5) & 0.414 & 0.449 & 0.447 & 0.417 & 0.753 & 0.480 & 3 & 3 & 3 & 3 & 2 & 4 \\ 
\end{tabular}
}
\end{center}
\end{table}

\subsection{Trend scenario}\label{sec:trend}

For the trend scenario (T) we consider trends in the mean and the standard deviation and model these using the function $h_\text{T}(t;\btheta)	\coloneqq \theta_0 + \theta_1 t \exp\left(\theta_2 t\right)$, where $\btheta \coloneqq (\theta_0, \theta_1, \theta_2)\in\R^{3}$.
Hence, for $t=1,\ldots,N$, we model the mean as
$M_\text{T}(t;\bmu) \coloneqq h_\text{T}(t;\bmu)$,
and the standard deviation as $S_\text{T}(t;\bsigma) \coloneqq h_\text{T}(t;\bsigma)$,
where $\bmu=(\mu_0,\mu_1,\mu_2)\in\R^{3}$ and $\bsigma=(\sigma_0,\sigma_1,\sigma_2)\in\R^{3}$, respectively.

\paragraph{Trend scenario (T)}
The subsequent equations define the data generating process (T0) and the five models (T1)--(T5), where we consider time series of length $N=200$.
\begin{alignat*}{4}
	& \mbox{(T0)}	&& \quad Y_t && \sim\calN(M_\text{T}(t;\bmu^0),[S_\text{T}(t;\bsigma^0)]^2)	&& \quad \mbox{(data generating process),}	\\
	\hline
	& \mbox{(T1)}	&& \quad X_t^1 && \sim\calN(M_\text{T}(t;\bmu^1),[S_\text{T}(t;\bsigma^1)]^2)	&& \quad \mbox{(true model),}	\\
	& \mbox{(T2)}	&& \quad X_t^2 && \sim\calN(M_\text{T}(t;\bmu^2),[S_\text{T}(t;\bsigma^2)]^2)	&& \quad \mbox{(wrong mean),}	\\
	& \mbox{(T3)}	&& \quad X_t^3 && \sim\calN(M_\text{T}(t;\bmu^3),[S_\text{T}(t;\bsigma^3)]^2)	&& \quad \mbox{(wrong variance),}	\\
	& \mbox{(T4)}	&& \quad X_t^4 && \sim\calN(M_\text{T}(t;\bmu^4),[S_\text{T}(t;\bsigma^4)]^2)	&& \quad \mbox{(wrong mean and lin. variance),}	\\
	& \mbox{(T5)}	&& \quad X_t^5 && \sim\calN(M_\text{T}(t;\bmu^5),[S_\text{T}(t;\bsigma^5)]^2)	&& \quad \mbox{(wrong mean and const. variance),}
	\end{alignat*}
\begin{alignat*}{4}
	& \mbox{(T0)}	&& \quad \bmu^0 = (0,1/3,2)/N,	&& \quad \bsigma^0 = (20,0.05,2)/N,	&& \quad \mbox{(data generating process),}	\\
	\hline
	& \mbox{(T1)}	&& \quad \bmu^1 = \bmu^0,	&& \quad \bsigma^1 = \bsigma^0,	&& \quad \mbox{(true model),}	\\
	& \mbox{(T2)}	&& \quad \bmu^2 = (0,1/3,1.9)/N,	&& \quad \bsigma^2 = \bsigma^0,	&& \quad \mbox{(wrong mean),}	\\
	& \mbox{(T3)}	&& \quad \bmu^3 = \bmu^0,	&& \quad \bsigma^3 = (20,0.0375,1.5)/N,	&& \quad \mbox{(wrong variance),}	\\
	& \mbox{(T4)}	&& \quad \bmu^4 = \bmu^2,	&& \quad \bsigma^4 = (20,0.05,0)/N,	&& \quad \mbox{(wrong mean and lin. variance),}	\\
	& \mbox{(T5)}	&& \quad \bmu^5 = \bmu^2,	&& \quad \bsigma^5 = (20,0,0)/N,	&& \quad \mbox{(wrong mean and const. variance).}
\end{alignat*}
Whereas the true model (T1) is equivalent to the data generating process (T0) in terms of its parametrization, models (T2)--(T5) differ from (T0) in at least one parameter. The mean of model (T2) is always too small while the variance is modeled correctly. Model (T3) models the exponentially increasing mean correctly with too small a variance. Models (T4) and (T5) are misspecified in both mean and variance, where the misspecification of the variance is less severe for model (T4).
Figure S.5 in the supplementary material illustrates one replication of the DGP (T0) and the corresponding selection of moving windows. Figure S.7 compares models (T1)--(T5) based on time series of theoretical/empirical CRPS. Here, the PELT algorithm estimates between four and eight changepoints in the simulated series resulting in a fixed window width ranging from $15$ to $51$.

\paragraph{Model rankings}
The resulting scores and the associated model rankings are shown in Table \ref{tab:Trank}. As the theoretical SE evaluates the mean of the model/prediction only, the competing models are here ranked in only two groups. Similarly as for the changepoint scenario above, the misspecified model with correct mean and too small a variance is ranked higher than the true model using the emprical moving scores and the same effect is observed for the CRPS. Again we observe erroneous model rankings under the PW and the ST approaches. 

\begin{table}[t]
\caption{\label{tab:Trank}\textbf{Trend scenario (T):} Comparison/ranking of models (T1)--(T5) based on average (empirical/theoretical) SE and CRPS. Average scores (left) and corresponding model rankings (right) are provided, distinguishing between theoretical (Th) and empirical scores computed using different approaches (OF, OV, DV, PW, ST). The rankings based on the theoretical scores are considered the true model rankings and can be used to judge the rankings given by the empirical scores.}
\begin{center}
\fbox{
\begin{tabular}{ll|c|ccccc|c|ccccc}
	&	&	\multicolumn{6}{c|}{average scores}	&	\multicolumn{6}{c}{model rank} \\ 
	& & Th & OF & OV & DV & PW & ST & Th & OF & OV & DV & PW & ST \\ 
  \hline
\multirow{5}{*}{\rotatebox[origin=c]{90}{SE}}
	& (T1) & 0.053 & 0.057 & 0.057 & 0.056 & 0.106 & 0.515 & 1 & 2 & 2 & 2 & 4 & 2 \\ 
  & (T2) & 0.059 & 0.063 & 0.063 & 0.064 & 0.112 & 0.518 & 3 & 5 & 5 & 5 & 5 & 5 \\ 
  & (T3) & 0.053 & 0.054 & 0.054 & 0.054 & 0.079 & 0.515 & 1 & 1 & 1 & 1 & 3 & 1 \\ 
  & (T4) & 0.059 & 0.059 & 0.060 & 0.062 & 0.075 & 0.518 & 3 & 4 & 4 & 4 & 2 & 4 \\ 
  & (T5) & 0.059 & 0.059 & 0.059 & 0.062 & 0.069 & 0.518 & 3 & 3 & 3 & 3 & 1 & 3 \\ 
	\hline
\multirow{5}{*}{\rotatebox[origin=c]{90}{CRPS}}
  & (T1) & 0.116 & 0.124 & 0.124 & 0.120 & 0.232 & 0.387 & 1 & 2 & 2 & 2 & 4 & 2 \\ 
  & (T2) & 0.121 & 0.128 & 0.128 & 0.126 & 0.237 & 0.388 & 3 & 3 & 3 & 3 & 5 & 5 \\ 
  & (T3) & 0.119 & 0.122 & 0.122 & 0.119 & 0.206 & 0.386 & 2 & 1 & 1 & 1 & 3 & 1 \\ 
  & (T4) & 0.131 & 0.128 & 0.129 & 0.127 & 0.201 & 0.388 & 4 & 4 & 4 & 4 & 2 & 3 \\ 
  & (T5) & 0.136 & 0.130 & 0.130 & 0.128 & 0.193 & 0.388 & 5 & 5 & 5 & 5 & 1 & 4 \\ 
\end{tabular}
}
\end{center}
\end{table}

\subsection{Periodicity scenario}\label{sec:periodicity}

For the periodicity scenario (P) we consider time series with periodically varying mean and standard deviation. To model the periodicity, we consider the function $h_\text{P}(t;\btheta)	\coloneqq \theta_0 + \theta_1 \sin\left(2\pi t \theta_2\right)$, where $\btheta \coloneqq (\theta_0, \theta_1, \theta_2)\in\R^{3}$.
Then, for $t=1,\ldots,N$, we model the mean as $M_\text{P}(t;\bmu) \coloneqq h_\text{P}(t;\bmu)$, and the standard deviation as $S_\text{P}(t;\bsigma) \coloneqq \exp\left(h_\text{P}(t;\bsigma)\right)$, with parameter vectors $\bmu=(\mu_0,\mu_1,\mu_2)\in\R^{3}$ and $\bsigma=(\sigma_0,\sigma_1,\sigma_2)\in\R^{3}$, respectively.

\paragraph{Periodicity scenario (P)}
The subsequent equations define the data generating process (P0) and the five models (P1)--(P5), where we consider time series of length $N=730$.
\begin{alignat*}{4}
	& \mbox{(P0)}	&& \quad	Y_t && \sim\calN(M_\text{P}(t;\bmu^0),[S_\text{P}(t;\bsigma^0)]^2)	&& 	\quad\mbox{(data generating process),}	\\
	\hline
	& \mbox{(P1)}	&& \quad	X_t^1 && \sim\calN(M_\text{P}(t;\bmu^1),[S_\text{P}(t;\bsigma^1)]^2)	&& 	\quad\mbox{(true model),}	\\
	& \mbox{(P2)}	&& \quad	X_t^2 && \sim\calN(M_\text{P}(t;\bmu^2),[S_\text{P}(t;\bsigma^2)]^2)	&& 	\quad\mbox{(wrong mean),}	\\
	& \mbox{(P3)}	&& \quad	X_t^3 && \sim\calN(M_\text{P}(t;\bmu^3),[S_\text{P}(t;\bsigma^3)]^2)	&& 	\quad\mbox{(wrong variance),}	\\
	& \mbox{(P4)}	&& \quad	X_t^4 && \sim\calN(M_\text{P}(t;\bmu^4),[S_\text{P}(t;\bsigma^4)]^2)	&& 	\quad\mbox{(wrong mean and variance),}	\\
	& \mbox{(P5)}	&& \quad	X_t^5 && \sim\calN(M_\text{P}(t;\bmu^5),[S_\text{P}(t;\bsigma^5)]^2)	&& 	\quad\mbox{(wrong mean and constant variance),}
\end{alignat*}
\pagebreak
\begin{alignat*}{4}
	& \mbox{(P0)}	&& \quad	\bmu^0 = (0,10,1/365),	&& \quad	\bsigma^0 = (0,-0.5,1/365),	&& \quad	\mbox{(data generating process),}	\\
	\hline
	& \mbox{(P1)}	&& \quad	\bmu^1 = \bmu^0,	&& \quad	\bsigma^1 = \bsigma^0,	&& \quad	\mbox{(true model),}	\\
	& \mbox{(P2)}	&& \quad	\bmu^2 = (0,9.5,1/365),	&& \quad	\bsigma^2 = \bsigma^0,	&& \quad	\mbox{(wrong mean),}	\\
	& \mbox{(P3)}	&& \quad	\bmu^3 = \bmu^0,	&& \quad	\bsigma^3 = (0,-0.25,1/365),	&& \quad	\mbox{(wrong variance),}	\\
	& \mbox{(P4)}	&& \quad	\bmu^4 = \bmu^2,	&& \quad	\bsigma^4 = \bsigma^3	&& \quad	\mbox{(wrong mean and variance),}	\\
	& \mbox{(P5)}	&& \quad	\bmu^5 = \bmu^2,	&& \quad	\bsigma^5 = (0,0,1/365),	&& \quad	\mbox{(wrong mean and constant variance).}
\end{alignat*}

Whereas the true model (P1) is equivalent to the data generating process (P0) in terms of its parametrization, models (P2)--(P5) differ from (P0) in at least one parameter. Model (P2) underestimates the magnitude of the mean oscillations, whereas it captures the variance oscillations correctly. Model (P3) captures the mean oscillations correctly, however, the oscillations in the variance are underestimated. Models (P4) and (P5) are misspecified both in terms of mean and variance. Model (P4) models some of the variation in the variance while Model (P5) assumes a constant variance. Figure S.6 in the supplementary material illustrates one replication of the DGP (P0) and the corresponding selection of moving windows. Figure S.8 compares models (P1)--(P5) based on time series of theoretical/empirical CRPS. For this scenario, the PELT algorithm estimates between $24$ and $34$ changepoints resulting in the fixed window width ranging from $13$ to $23$.

\paragraph{Model rankings}
Table \ref{tab:Prank} shows scores and model rankings for models (P1)--(P5) under the different evaluation approaches. Here, the empirical rankings for the CRPS calculated based on the OF, OV and DV approach are identical to the corresponding theoretical rankings. As before, the SE is unable to differentiate between models with identical mean value structure and the empirical approaches fail to recognize the true model compared to a model with correct mean structure and too small spread. Again, the PW approach yields wrong model rankings.

\begin{table}[t]
\caption{\label{tab:Prank}\textbf{Periodicity scenario (P):} Comparison/ranking of models (P1)--(P5) based on average (empirical/theoretical) SE and CRPS. Average scores (left) and corresponding model rankings (right) are provided, distinguishing between theoretical (Th) and empirical scores computed using different approaches (OF, OV, DV, PW). The rankings based on the theoretical scores are considered the true model rankings and can be used to judge the rankings given by the empirical scores.}
\begin{center}
\fbox{
\begin{tabular}{ll|c|cccc|c|cccc}
	&	&	\multicolumn{5}{c|}{average scores}	&	\multicolumn{5}{c}{model rank} \\ 
	& & Th & OF & OV & DV & PW & Th & OF & OV & DV & PW \\ 
  \hline
\multirow{5}{*}{\rotatebox[origin=c]{90}{SE}}
	& (P1) & 1.266 & 1.341 & 1.361 & 1.427 & 2.531 & 1 & 2 & 2 & 2 & 4 \\ 
  & (P2) & 1.391 & 1.485 & 1.589 & 1.574 & 2.656 & 3 & 5 & 5 & 5 & 5 \\ 
  & (P3) & 1.266 & 1.330 & 1.357 & 1.424 & 2.328 & 1 & 1 & 1 & 1 & 1 \\ 
  & (P4) & 1.391 & 1.474 & 1.585 & 1.571 & 2.453 & 3 & 4 & 3 & 3 & 3 \\ 
  & (P5) & 1.391 & 1.471 & 1.586 & 1.572 & 2.390 & 3 & 3 & 4 & 4 & 2 \\ 
	\hline
\multirow{5}{*}{\rotatebox[origin=c]{90}{CRPS}}
  & (P1) & 0.600 & 0.643 & 0.645 & 0.657 & 1.200 & 1 & 1 & 1 & 1 & 2 \\ 
  & (P2) & 0.638 & 0.683 & 0.699 & 0.694 & 1.238 & 3 & 3 & 3 & 3 & 5 \\ 
  & (P3) & 0.605 & 0.646 & 0.646 & 0.659 & 1.178 & 2 & 2 & 2 & 2 & 1 \\ 
  & (P4) & 0.641 & 0.684 & 0.700 & 0.695 & 1.214 & 4 & 4 & 4 & 4 & 3 \\ 
  & (P5) & 0.652 & 0.694 & 0.706 & 0.702 & 1.216 & 5 & 5 & 5 & 5 & 4 \\ 
\end{tabular}
}
\end{center}
\end{table}

\section[Case study: Evaluation of Regional Climate Models]{Case study: Evaluation of Regional Climate Models}\label{sec:expl}

In an application of the presented methodology, we evaluate four selected \emph{Regional Climate Models (RCMs)} from the \emph{ENSEMBLES project} \citep[][]{linden09}. The purpose of RCMs is to dynamically downscale gridded climate data from global models with a coarse spatial resolution (say $\sim200$--$300$ km) to obtain climate simulations of a finer resolution ($\sim25$--$50$ km) for a specific region. As a part of the ENSEMBLES project, ensembles of RCM simulations from $16$ different (European) climate research institutes were compiled. In an initial $40$-year experiment ($1961$--$2000$) covering Europe, the \emph{RCMs were driven by the ERA-40 reanalysis data set}. This experiment was used to evaluate the RCMs (see also Section \ref{sec:intro}). Furthermore, \emph{RCM experiments driven by Global Circulation Model (GCM) output} were conducted to create an ensemble of regional climate change projections for Europe. Most of these RCM simulations cover the period $1951$--$2050$, and are available on a daily temporal resolution. For our case study we consider only the output variable \emph{2 meter temperature (daily mean temperature)}. For the control period $1961$--$1990$ and a spatial resolution of $25$ kilometers, we compare four RCMs:
\begin{itemize}
	\item \emph{DMI-HIRHAM} of the Danish Meteorological Institute (\emph{DMI}),
	\item \emph{KNMI-RACMO2} of the Royal Netherlands Meteorological Institute (\emph{KNMI}),
	\item \emph{MPI-M-REMO} of the Max-Planck-Institute for Meteorology (\emph{MPI}), and
	\item \emph{SMHIRCA} of the Swedish Meteorological and Hydrological Institute (\emph{SMHI}).
\end{itemize}
In the following, we refer to the different models by the acronym of the corresponding institute. We evaluate the models for both ERA-40 and GCM boundary conditions. All four RCM simulations we consider were driven by the \emph{ECHAM5} GCM. To differentiate between the two boundary conditions, we use the acronyms \emph{ERA-40} and \emph{ECHAM5}.

The RCM model output are evaluated against the E-OBS observational data product \citep[version 13.1,][]{haylock08}. This data product is a gridded data set interpolated from station observations. It covers the European continent (land only) and is available for different grids and spatial resolutions. For our evaluation we use the gridded \emph{daily mean temperature} observations provided in version 13.1 of the data set. Our evaluation is based on all days in the period $1961$-$1990$ at a total of $10937$ grid points. 

\subsection{Evaluation approaches}\label{sec:approaches}
To evaluate and compare the different RCMs, we calculate moving CRPS based on OF-windows (see Equation \eqref{eq:OFwindows}), OV-windows (see Equation \eqref{eq:OVwindows}), and DV-windows (see Equation \eqref{eq:DVwindows}). For comparison, we also compute the point-wise (PW) scores for comparison. Recall that the point-wise CRPS is equivalent to the absolute error.

\subsection{Overall model assessment}\label{sec:overall}
Table \ref{tab:osd} provides overall (spatial and temporal) averages of moving CRPS and point-wise CRPS. As expected, the ERA-40 driven models achieve better average scores than the GCM driven models. While the CRPS calculations yield identical model rankings for all three types of moving windows (OF, OV, DV), the point-wise scores yield a different ranking. As we learned in our simulation study in Section \ref{sec:sim}, point-wise scores might lead to a spurious model ranking, since they do not account for higher order structures of the observed/modeled phenomenon. The results of Table \ref{tab:osd} corroborate these findings. Moving scores provide a more holistic model assessment and hence should be preferred over a point-wise evaluation. It seems, that for the application at hand the type of the moving window is not crucial.

\begin{table}[t]
\caption{Overall (spatial and temporal) averages of moving CRPS and point-wise (PW) CRPS for the DMI, KNMI, MPI and SMHI models driven by ECHAM5 (top) and ERA-40 (bottom). The relative score-based rank of each model (distinguishing between ECHAM5 and ERA-40 boundary conditions) is given in brackets.}
	\label{tab:osd}
\begin{center}
\fbox{
\begin{tabular}{ll|rrrrrrrr}
ECHAM5 &  & \multicolumn{2}{c}{DMI} & \multicolumn{2}{c}{KNMI} & \multicolumn{2}{c}{MPI} & \multicolumn{2}{c}{SMHI} \\ 
  \hline
& OF & 2.73 & (\nth{4}) & 2.42 & (\nth{1}) & 2.49 & (\nth{3}) & 2.46 & (\nth{2}) \\ 
CRPS & OV & 2.70 & (\nth{4}) & 2.41 & (\nth{1}) & 2.48 & (\nth{3}) & 2.45 & (\nth{2}) \\ 
& DV & 2.55 & (\nth{4}) & 2.26 & (\nth{1}) & 2.32 & (\nth{3}) & 2.29 & (\nth{2}) \\ 
  \hline
& PW & 4.57 & (\nth{4}) & 4.17 & (\nth{3}) & 4.07 & (\nth{2}) & 3.99 & (\nth{1}) \\ 
  \hline
	\hline
ERA-40 &  & \multicolumn{2}{c}{DMI} & \multicolumn{2}{c}{KNMI} & \multicolumn{2}{c}{MPI} & \multicolumn{2}{c}{SMHI} \\ 
  \hline
& OF & 1.95 & (\nth{1}) & 1.98 & (\nth{2}) & 2.07 & (\nth{4}) & 1.99 & (\nth{3}) \\ 
CRPS & OV & 1.94 & (\nth{1}) & 1.96 & (\nth{2}) & 2.05 & (\nth{4}) & 1.98 & (\nth{3}) \\ 
& DV & 1.77 & (\nth{1}) & 1.79 & (\nth{2}) & 1.89 & (\nth{4}) & 1.81 & (\nth{3}) \\ 
  \hline
& PW & 2.02 & (\nth{2}) & 2.11 & (\nth{3}) & 2.24 & (\nth{4}) & 1.99 & (\nth{1}) \\ 
\end{tabular}
}
\end{center}
\end{table}

\subsection{Temporal evaluation}\label{sec:temp}

\begin{figure}[hbp]
	\centering
		\includegraphics[width=1.0\textwidth,trim={0 0.6cm 0 0},clip]{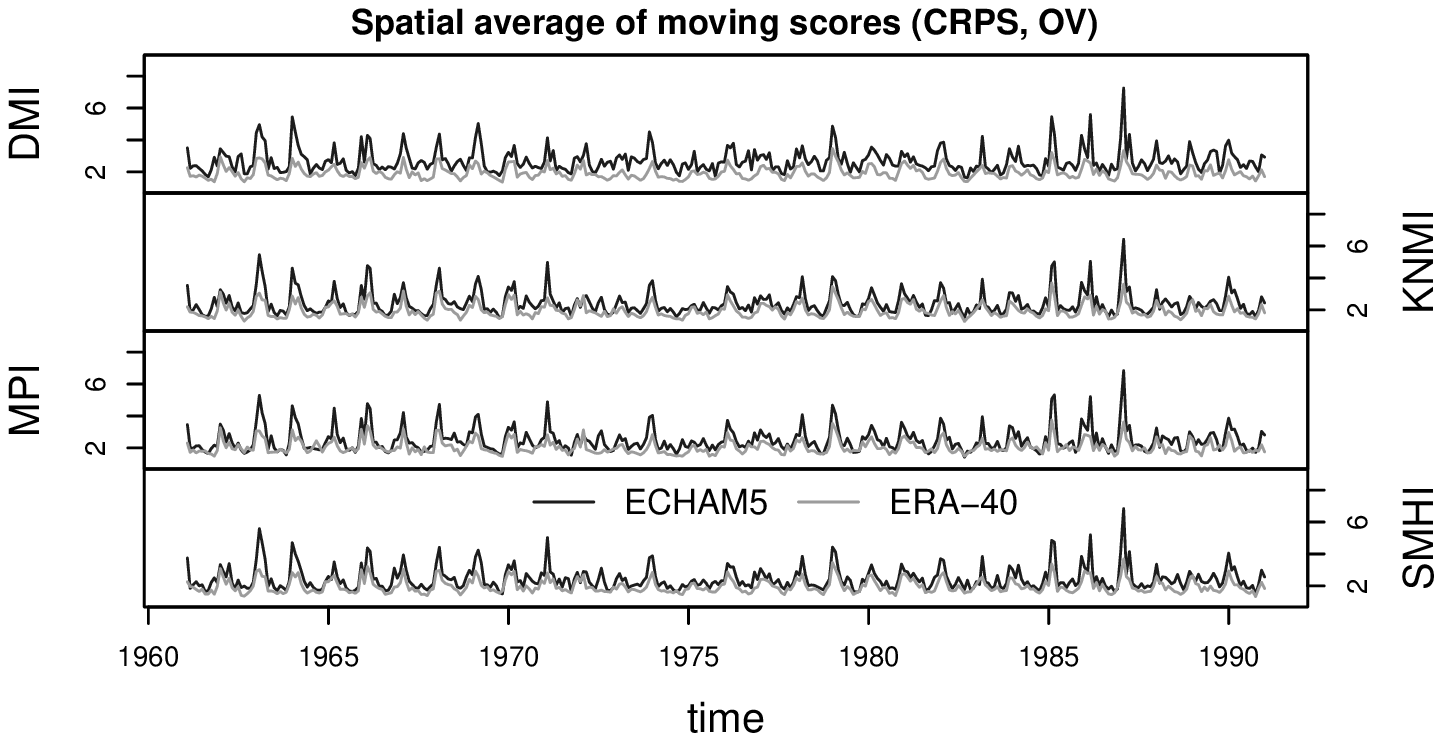}
		\includegraphics[width=1.0\textwidth,trim={0 0.6cm 0 0},clip]{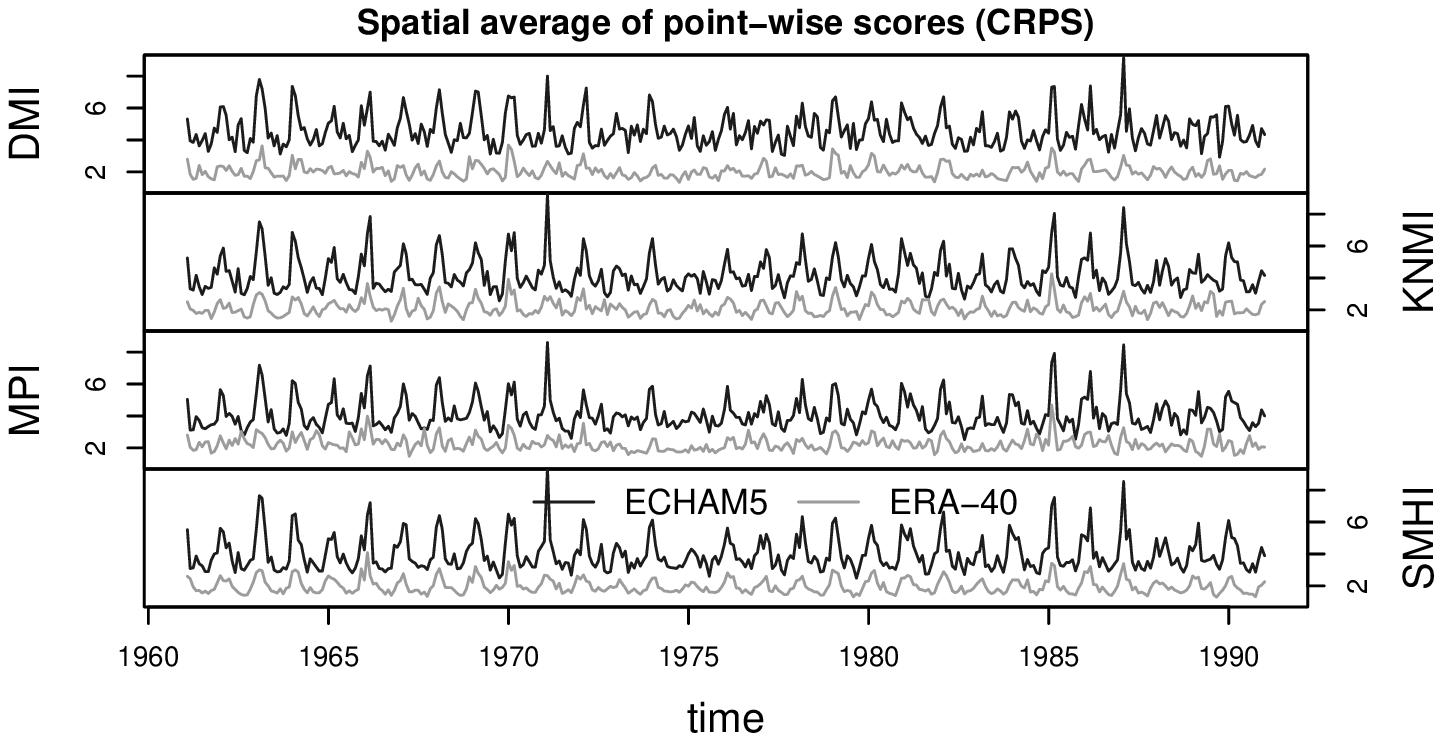}
	\caption{Moving CRPS based on OV-windows (top) and point-wise CRPS (bottom) aggregated over months and the entire study area for the DMI, KNMI, MPI and SMHI models driven by ECHAM5 (black) and ERA-40 (gray).}
	\label{fig:csOVts}
\end{figure}

To assess the model accuracy over time we consider spatial average scores over the entire study area, see Figure \ref{fig:csOVts}. For illustration purposes, we further smooth the averaged daily series by computing monthly averages. Figure \ref{fig:csOVts} compares the resulting time series for the moving CRPS with OV-windows, and for the point-wise CRPS. The corresponding figures for OF- and DV-windows can be found in the supplementary material (Figure S.9).

All four models show pronounced seasonal oscillations in the scores, indicating a better performance in the summer and a worse performance in the winter. The osciallations are less pronounced for the ERA-40 driven models than for the ECHAM5 driven models. From that we conclude that the driving GCM (here ECHAM5) is not able to accurately capture the (regional) seasonal temperature dynamics. While for the moving CRPS the differences between the different models and the different boundary conditions (ECHAM5 and ERA-40) appear to be rather small, the point-wise evaluation indicates larger differences. This is due to the fact that the ECHAM5 boundary conditions are simulations from a GCM and, as such, are not designed to capture the weather in a particular month or a season, while the ERA-40 boundary conditions are a reanalysis based on observations which should capture the changes in weather. 

Focusing on the moving CRPS, we observe time periods of weaker (e.g. $1985$--$1987$) and of better (e.g. $1974$--$1977$) model performance. Moreover, we find that for the DMI model the discrepancy between the two different boundary conditions is more obvious compared to the other models. While the ERA-40 driven DMI model apparently performs best in comparison to the other models, the ECHAM5 driven DMI model performs worst (see also Table \ref{tab:osd}).

\subsection{Spatial evaluation}\label{sec:spat}
To judge the model performance for different locations across Europe we look at the temporal averages of the scores. The temporal averages corresponding to the moving CRPS based on OV-windows and the point-wise CRPS are depicted in Figure \ref{fig:csOVmaps}. The corresponding figures for OF- and DV-windows are provided in the supplementary material (Figure S.10).

\begin{figure}[hbp]
	\centering
		\includegraphics[width=1.0\textwidth,trim={0 0.5cm 0 0},clip]{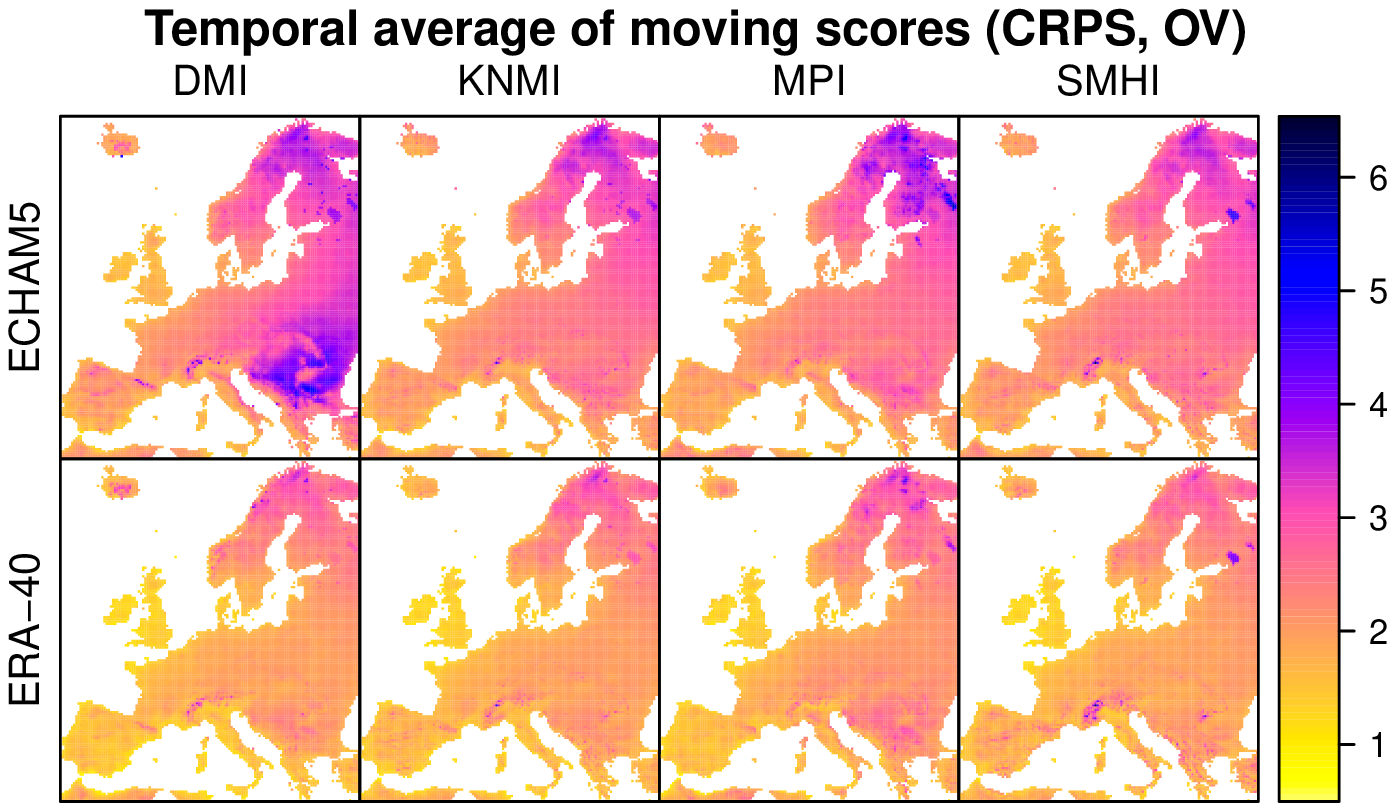}
		\includegraphics[width=1.0\textwidth,trim={0 0.5cm 0 0},clip]{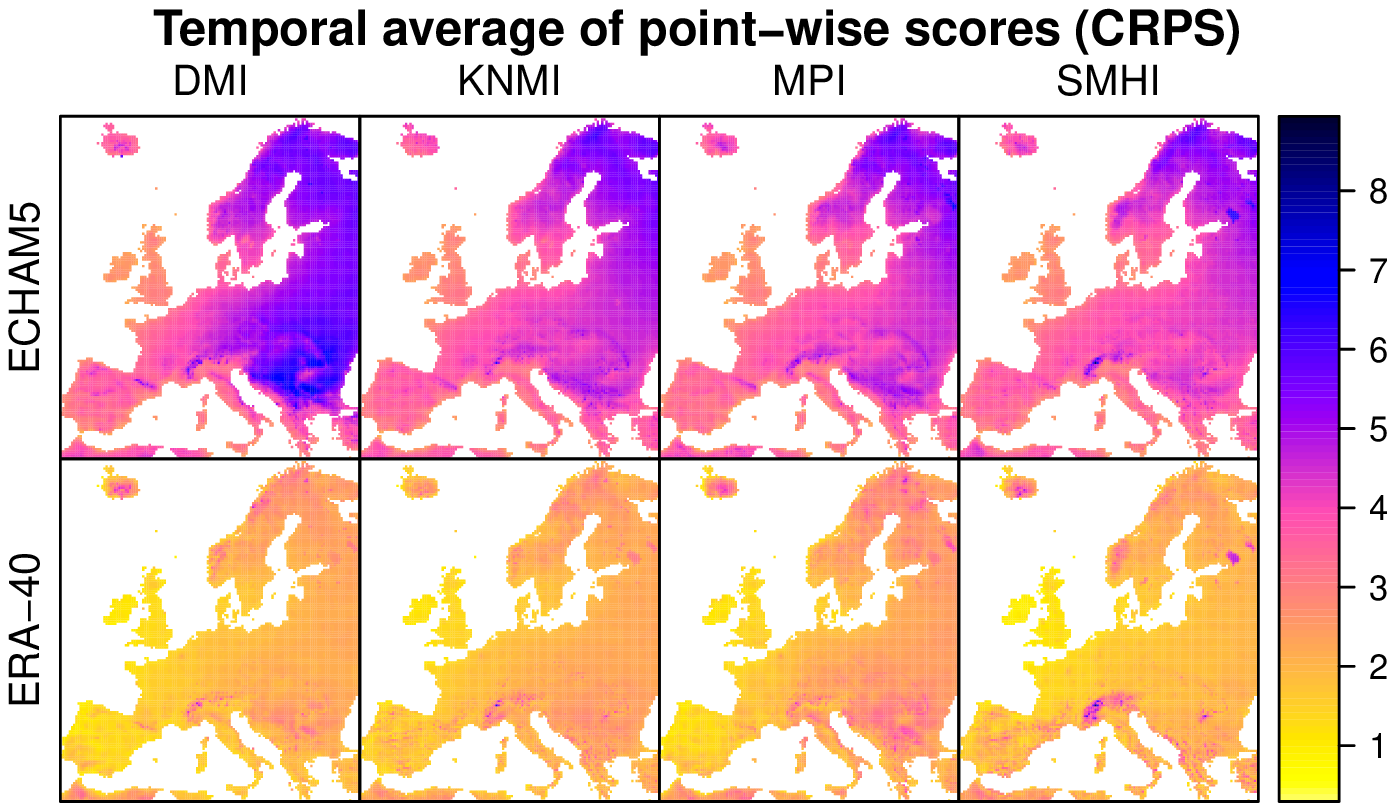}
	\caption{Maps of Europe showing temporal averages ($1961$--$1990$) of moving CRPS based on OV-windows (top) and point-wise CRPS (bottom), for the DMI, KNMI, MPI and SMHI models (ECHAM5/ERA-40).}
	\label{fig:csOVmaps}
\end{figure}

Again the distinction between the ERA-40 and the ECHAM5 boundary conditions is more pronounced for the point-wise score which does not account for higher order structures in the temperature series. In particular, a point-wise evaluation does not account for the associated uncertainty which may vary across space. Among the ECHAM5 driven models, the DMI model performs worst, especially regarding the continental climate in eastern Europe. The MPI model under ECHAM5 performs badly in north-eastern Europe where lakes dominate the landscape. Figure \ref{fig:csOVmaps} indicates that all the models have difficulties in modeling the temperature over large water bodies (e.g. lakes Ladoga and Onega in Russia). This is identified to a lesser degree by the point-wise score. Furthermore, we observe that most models have problems for mountainous areas (e.g. Alps, Pyrenees, Carpathians). To show for each pixel of the study area which model has the lowest (best) average score, we provide Figure S.11 in the supplementary material.

\subsection{Evaluation based on linear trends}\label{sec:ltrend}

For comparison, we now consider one of the evaluation approaches undertaken in the literature. Following the approach of \citet[][]{lorenz10}, we compare linear temperature trends in the RCM output and the E-OBS reference data.

To quantify the trends, we first aggregate the daily time series to yearly time series $y_t$, $t=1961,\ldots,1990$, by computing annual means. We then assume a linear regression model $y_t = \alpha + \beta t + \veps_t$, with intercept $\alpha$, linear trend parameter $\beta$ and residuals $\veps_t$ ($t=1961,\ldots,1990$). The parameter estimates are obtained using least-squares estimation separately for all $n=10937$ pixels/locations $i=1,\ldots,n$ of the study area and each data set. Subsequently, we distinguish between estimates $\wh\beta_i^\text{mod}$ corresponding to one of the RCM outputs and estimates $\wh\beta_i^\text{ref}$ corresponding to the reference data set. To compare and rank the different RCMs, we finally compute absolute trend errors of the form $\left|\wh\beta_i^\text{mod}-\wh\beta_i^\text{ref}\right|$, $i=1,\ldots,n$.

To summarize the results of the linear trend-based evaluation, we consider spatial averages $1/n\sum_{i=1}^n\left|\wh\beta_i^\text{mod}-\wh\beta_i^\text{ref}\right|$ of the absolute trend errors, see Table \ref{tab:tdiff}. For the ECHAM5 boundary conditions, the DMI model is considered best. For ERA-40 boundary conditions, the models are ranked in the order SMHI (best), KNMI, DMI and MPI (worst). These results differ considerably from the score-based evaluation in Table \ref{tab:osd}. Here, direct comparison of the DMI model for the two different boundary conditions further suggests that the model driven by ECHAM5 performs better. To see for each pixel of the study area which of the four RCMs has the smallest absolute trend error, we provide Figure S.12 in the supplementary material.

\begin{table}[t]
\caption{Average absolute decadal trend errors aggregated over all grid points within the study area.}
	\label{tab:tdiff}
\centering
\fbox{
\begin{tabular}{lllll}
 & DMI & KNMI & MPI & SMHI \\ 
  \hline
	ECHAM5 & 0.14 & 0.28 & 0.28 & 0.25 \\ 
  ERA-40 & 0.15 & 0.14 & 0.20 & 0.13 \\ 
\end{tabular}
}
\end{table}


\section[Conclusions and outlook]{Conclusions and outlook}\label{sec:conc}

We propose methodology for the evaluation of time series models/predictions under the presence of non-stationarity. Our approach utilizes proper scores on moving time windows where stationarity is assumed, in order to provide a fair and holistic comparison of models/predictions. A simulation study explores the moving score technique under the presence of changepoints, trends and periodicity. A case study, comparing model output from regional climate models, illustrates the utility of the technique for practical applications. We summarize our results and conclusions in the following.

Being based on proper scores, our approach allows for a \emph{fair} comparative model/prediction evaluation.
Evaluation based on moving windows instead of a point-wise comparison accounts for \emph{higher order structures} of the modeled/predicted phenomenon. Compared to most other evaluation approaches used in practice our evaluation method does not require a separate consideration of \emph{different seasons} when considering sub-seasonal evaluation, or to make a (subjective) decision on the importance/weight of \emph{different features of the model/prediction}. To base the \emph{choice of the moving windows} on (detected) changepoints is meaningful, since changepoints divide a time series into stationary segments.

Note that the sequential comparison of individual observations within a time window against an empirical distribution of the model output from the same time window using a proper scoring rule is equivalent to a direct comparison of the model distribution and the corresponding observation distribution using an associated score divergence \citep{thorarinsdottir13}. In a multi-model comparison, the proper scoring rule and its score divergence will yield the same model rankings. However, the application of the divergence function in our setting requires some care, as overlapping moving windows and windows of varying sizes may result in aggregated values that include unequal weighting of the observations. Such weighting may destroy the propriety of the metric \citep{gneiting2011}.    

Depending on the phenomenon of interest, certain \emph{types of moving windows} might be more adequate than others. For instance, disjoint windows with varying width (DV) for fixed changepoints, overlapping windows with fixed (OF) or varying width (OV) under the presence of trends or periodicity, etc. Similar as in the simulated examples, our case study showed \emph{identical overall model rankings} based on moving CRPS for different window selection approaches. For the application of the moving score methodology to \emph{continuous outcomes} utilization of the CRPS is preferred over the SE score as it evaluates the full predictive distribution rather than the predictive mean only. In general, the simulation study showed that moving scores are able to \emph{approximate their theoretical counterparts} considerably well, are \emph{better suited than point-wise (PW) scores or scores under a stationarity assumption (ST)} to yield an adequate (true) model/prediction ranking and yield the \emph{correct model ranking} in most cases (see the trend scenario (Section \ref{sec:trend}) for an exception). 

An extension of the work presented here to categorical outcomes is of interest. For example, one might want to judge models which differentiate only if a certain (extreme) event occurs or not. Further extensions of our approach might concern the judgment of model accuracy in a spatial context, where we can think of the moving window as a spatial neighborhood considering either a fixed number of spatial neighbors or all locations within a certain radius.



\section*{Acknowledgements}

Tobias Erhardt is supported by Deutsche Forschungsgemeinschaft (DFG) through the TUM International Graduate School of Science and Engineering (IGSSE). Claudia Czado is supported by the DFG (grant CZ86/4-1). Thordis L. Thorarinsdottir is supported by the Research Council of Norway through grant number 243953 ``Physical and Statistical Analysis of Climate Extremes in Large Datasets'' (ClimateXL).  
The ENSEMBLES and the E-OBS data used in this work were funded by the EU FP6 Integrated Project ENSEMBLES (Contract number 505539, \texttt{http://ensembles-eu.metoffice.com}) which we gratefully acknowledge. Moreover, we acknowledge the data providers of the E-OBS dataset in the ECA\&D project (\texttt{http://www.ecad.eu}).
The numerical computations were performed on a Linux cluster supported by DFG grant INST 95/919-1 FUGG.

\bibliographystyle{apalike}
\bibliography{literature}

\label{lastpage}

\end{document}